\documentclass[structabstract]{aa}

\usepackage{graphicx}
\usepackage{txfonts}
\usepackage{epsfig}

\begin{document}
   \title{The helium star donor channel for the progenitors of type Ia supernovae with different metallicities}

   \author{B. Wang
          \inst{1,2,3}
          \and
          Z. Han \inst{1,2}
          }


   \institute{National Astronomical Observatories/Yunnan Observatory,
              the Chinese Academy of Sciences, Kunming 650011, China\\
              \email{wangbo@ynao.ac.cn, zhanwenhan@ynao.ac.cn}
              \and
              Key Laboratory for the Structure and Evolution of Celestial Objects, the Chinese Academy of Sciences, Kunming 650011, China
              \and
              Graduate University of the Chinese Academy of Sciences, Beijing 100049, China}

   \date{Received ; accepted}


  \abstract
{The nature of type Ia supernovae (SNe Ia) is still unclear.
Metallicities may have an important effect on their properties.}
{In this paper, we study the He star donor channel towards SNe Ia
comprehensively and systematically at various metallicities.}
{Employing Eggleton's stellar evolution code with the optically
thick wind assumption, we calculated about 10 000 WD + He star
systems and obtained SN Ia production regions of the He star donor
channel with metallicities $Z=0.03$, 0.02, 0.004 and 0.0001.
According to a detailed binary population synthesis approach, we
also obtained SN Ia birthrates at various metallicities.}
{Our study shows that both the initial mass of the He donor star and
the initial orbital period for SNe Ia increase with metallicity,
while the minimum initial mass of the carbon--oxygen white dwarfs
producing SNe Ia decreases with metallicity. For a constant
star-formation galaxy, SN Ia birthrates increase with metallicity.
If a single starburst is assumed, SNe Ia occur systemically earlier
and the peak value of the birthrate is larger for a high $Z$, and
the He star donor channel with different metallicities can produce
the young SNe Ia with delay times $\sim$45$-$220\,Myr.}
  {}


\titlerunning{Progenitors of SNe Ia with different metallicities}
\authorrunning{B. Wang \& Z. Han}

   \maketitle

%

\section{Introduction} \label{1. Introduction}
Type Ia supernova (SN Ia) explosions are among the most energetic
events observed in the Universe. They appear to be good cosmological
distance indicators and have been applied successfully in
determining cosmological parameters (e.g., $\Omega_{M}$ and
$\Omega_{\Lambda}$; Riess et al. 1998; Perlmutter et al. 1999). The
Phillips relation (a linear relation between the absolute magnitude
of SNe Ia and the magnitude difference from maximum to 15\,d after
\textit{B} maximum light) is adopted when SNe Ia are used as
distance indicators (Phillips 1993), which is based on the SN Ia
sample of the low red-shift Universe ($z<0.05$) and assumed to be
valid at high red-shift. This assumption is precarious since there
is still no agreement on the nature of their progenitors
(Hillebrandt \& Niemeyer 2000; Podsiadlowski et al. 2008; Wang et
al. 2008a; Gilfanov \& Bogd$\acute{\rm a}$n 2010; Mennekens et al.
2010). If the properties of SNe Ia evolve with red-shift, the
results for cosmology might be different. Since metallicity may
represent red-shift to some extent, it will be a good method to
study the properties of SNe Ia at various red-shift by finding the
correlation between their properties and metallicity. In addition,
some numerical and synthetical results show that metallicity may
have an effect on the final amount of nickel-56, and thus the
maximum luminosity of SNe Ia (Timmes et al. 2003; Podsiadlowski et
al. 2006; Podsiadlowski 2010). There is also some other evidence of
the correlation between the properties of SNe Ia and metallicity
from observations (e.g., Branch \& Bergh 1993; Hamuy et al. 1996;
Wang et al. 1997; Cappellaro et al. 1997; Shanks et al. 2002).

It is generally believed that SNe Ia are thermonuclear explosions of
carbon--oxygen white dwarfs (CO WDs) in binaries (Nomoto et al.
1997; Livio 2000). Over the past few decades, two families of SN Ia
progenitor models have been proposed, i.e., the double-degenerate
(DD) and single-degenerate (SD) models. It is suggested that the DD
model, which involves the merger of two CO WDs (Iben \& Tutukov
1984; Webbink 1984; Han 1998), likely leads to an accretion-induced
collapse rather than to an SN Ia (Nomoto \& Iben 1985; Timmes et al.
1994). For the SD model, the companion could be a main-sequence (MS)
star or a slightly evolved star (WD + MS channel), or a red-giant
star (WD + RG channel) (e.g., Hachisu et al. 1996, 1999a,b; Li $\&$
van den Heuvel 1997; Yungelson \& Livio 1998; Langer et al. 2000;
Fedorova et al. 2004; Han $\&$ Podsiadlowski 2004, 2006; Chen $\&$
Li 2007; L\"{u} et al. 2009; Meng \& Yang 2010; Wang et al. 2010a,b;
Wang \& Han 2010a). Note that some recent observations have
indirectly implied that at least some SNe Ia can be produced by a
variety of different progenitor systems (e.g., Hansen 2003;
Ruiz-Lapuente et al. 2004; Patat et al. 2007; Voss \& Nelemans 2008;
Wang et al. 2008b; Justham et al. 2009).

Yoon $\&$ Langer (2003) followed the evolution of a CO WD + He star
system with a $1.0\,M_{\odot}$ CO WD and a $1.6\,M_{\odot}$ He star
in a 0.124\,d orbit. In this binary, the WD accretes He from the He
star and grows in mass to the Chandrasekhar (Ch) mass. Recently,
Wang et al. (2009a) studied the He star donor channel of SNe Ia, in
which a CO WD accretes material from an He MS star or a slightly
evolved He subgiant to increase its mass to the Ch mass. The study
shows the parameter space for the progenitors of SNe Ia with
$Z=0.02$. By using a detailed binary population synthesis (BPS)
approach, Wang et al. (2009b) found that the Galactic SN Ia
birthrate from this channel is $\sim$$0.3\times 10^{-3}\ {\rm
yr}^{-1}$ and that this channel can produce the SNe Ia with short
delay times ($\sim$45$-$140\,Myr) from the star formation to SN
explosion (see also Wang \& Han 2010b). Considering that not all SNe
Ia are found in the solar metallicity environment ($Z=0.02$), we
will pay attention to the correlation between the properties of SNe
Ia and metallicities in this paper.

The purpose of this paper is to study the He star donor channel
towards SNe Ia comprehensively and systematically at various
metallicities, and then to determine the parameter space for SNe Ia,
which can be used in BPS studies. In Sect. 2, we describe the
numerical code for the binary evolution calculations and the grid of
the binary models. The binary evolutionary results are shown in
Sect. 3. We describe the BPS method in Sect. 4 and present the BPS
results in Sect. 5. Finally, a discussion is given in Sect. 6.

\section{Binary evolution calculations}\label{BINARY EVOLUTION CALCULATIONS}
In WD + He star systems, the He star fills its Roche lobe at He MS
or He subgiant stage, and then the mass transfer begins. The He star
transfers some of its material onto the surface of the WD, which
increases the mass of the WD as a consequence. We assume that, if
the WD grows to 1.378\,$M_{\odot}$, it explodes as an SN Ia.

\subsection{Stellar evolution code}
We use Eggleton's stellar evolution code (Eggleton 1971, 1972, 1973)
to calculate the evolution of the WD + He star systems. The code has
been updated with the latest input physics over the past four
decades (Han et al. 1994; Pols et al. 1995, 1998). Roche lobe
overflow (RLOF) is treated within the code described by Han et al.
(2000). We set the ratio of mixing length to local pressure scale
height, $\alpha=l/H_{\rm p}$, to be 2.0.  The opacity tables are
compiled by Chen \& Tout (2007) from Iglesias \& Rogers (1996) and
Alexander \& Ferguson (1994). Four metallicities are adopted in this
study (i.e., $Z=0.03$, 0.02, 0.004 and 0.0001). Orbital angular
momentum loss due to gravitational wave radiation (GWR) is also
included.

\subsection{WD mass growth}
Instead of solving stellar structure equations of a WD, we use an
optically thick wind model (Kato \& Hachisu 1994; Hachisu et al.
1996) and adopt the prescription of Kato \& Hachisu (2004, KH04) for
the mass accumulation efficiency of He-shell flashes onto the WD. If
the mass transfer rate, $|\dot M_2|$, is above a critical rate,
$\dot M_{\rm cr}$, we assume that He burns steadily on the surface
of the WD and that the He-rich material is converted into C and O at
a rate $\dot M_{\rm cr}$. The unprocessed matter is lost from the
system, presumably in the form of the optically thick wind at a mass
loss rate $\dot M_{\rm wind}=|\dot M_2| - \dot M_{\rm cr}$. Based on
the opacity from Iglesias \& Rogers (1996), the optically thick wind
is sensitive to Fe abundance, and it is likely that the wind does
not work when $Z$ is lower than a certain value 0.002 (Kobayashi et
al. 1998). Thus, there should be an obvious low-metallicity
threshold for SNe Ia in comparison with SN II. However, this
metallicity threshold has not been found (Prieto et al. 2008).
Considering the uncertainties in the opacities, we assume that the
optically thick wind is still valid for a low metallicity $Z=0.0001$
(see also Meng et al. 2009).

The critical mass transfer rate is
\begin{equation}
\dot M_{\rm cr}=7.2\times 10^{-6}\,(M_{\rm
WD}/M_{\odot}-0.6)\,M_{\odot}\,\rm yr^{-1},
\end{equation}
based on WD models computed with constant mass accretion rates
(Nomoto 1982). Similar to the work of Wang et al. (2009a), following
assumptions are adopted when $|\dot M_2|$ is smaller than $\dot
M_{\rm cr}$. (1) If $|\dot M_2|$ is less than $\dot M_{\rm cr}$ but
higher than the minimum accretion rate of stable He-shell burning,
$\dot M_{\rm st}$ (KH04), it is assumed that the He-shell burning is
stable and that there is no mass loss. (2) If $|\dot M_2|$ is less
than $\dot M_{\rm st}$ but higher than the minimum accretion rate of
weak He-shell flashes, $\dot M_{\rm
low}=4.0\times10^{-8}\,M_{\odot}\,\rm yr^{-1}$ (Woosley et al.
1986), He-shell flashes occur and a part of the envelope mass is
assumed to be blown off from the surface of the WD. The mass growth
rate of WDs in this case is linearly interpolated from a grid
computed by KH04, where a wide range of WD masses and accretion
rates was calculated in the He-shell flashes. (3) If $|\dot M_2|$ is
lower than $\dot M_{\rm low}$, the He-shell flashes will be so
strong that no mass can be accumulated onto the WD.

We define the mass growth rate of the CO WD, $\dot{M}_{\rm CO}$, as
 \begin{equation}
 \dot{M}_{\rm CO}=\eta _{\rm He}|\dot{M}_{\rm 2}|,
  \end{equation}
where $\eta _{\rm He}$ is the mass accumulation efficiency for
He-shell burning. According to the assumptions above, the values of
$\eta _{\rm He}$ are:
\begin{equation}
\eta_{\rm He}= \left\{ \begin{array}{l@{\quad,\quad}l}
\dot M_{\rm cr}\over |\dot M_2| &  |\dot{M_2}|>\dot{M}_{\rm cr},\strut\\
1\, &  \dot{M}_{\rm cr}\ge |\dot{M_2}|\ge \dot{M}_{\rm st},\strut\\
\eta'_{\rm He}\, &  \dot{M}_{\rm st}> |\dot{M_2}|\ge \dot{M}_{\rm low},\strut\\
0 &  |\dot{M_2}|< \dot{M}_{\rm low}.\strut\\
\end{array} \right.
\end{equation}

\subsection{Grid calculations}
We incorporate the prescriptions above into Eggleton's stellar
evolution code and follow the evolutions of the WD + He star
systems. The mass lost from these systems is assumed to take away
specific orbital angular momentum of the accreting WD. We calculated
about 10 000 WD + He star systems, and obtained a large, dense model
grid with four different metallicities. The initial mass of the He
donor stars, $M_{\rm 2}^{\rm i}$, ranges from $0.8\,M_{\odot}$ to
$3.3\,M_{\odot}$; the initial mass of the CO WDs, $M_{\rm WD}^{\rm
i}$, is from $0.86\,M_{\odot}$ to $1.20\,M_{\odot}$; the initial
orbital period of the binary systems, $P^{\rm i}$, changes from the
minimum value, at which an He zero-age MS (ZAMS) star would fill its
Roche lobe, to $\sim$400\,d, where the He star fills its Roche lobe
at the end of the Hertzsprung gap.

\section{Binary evolution results}
\subsection{An example of binary evolution calculations}
\begin{figure*}
\centerline{\epsfig{file=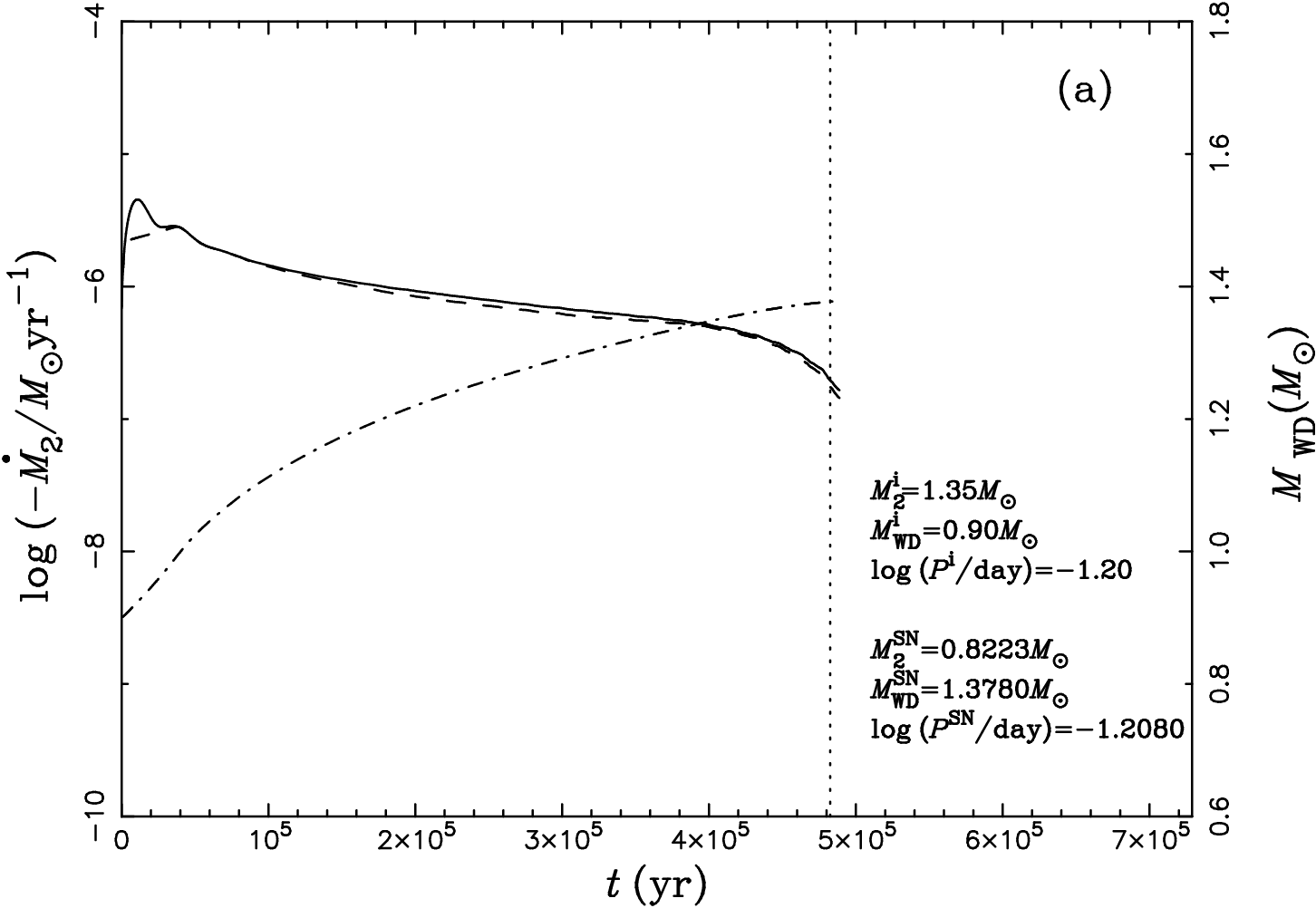,angle=270,width=8cm}\ \
\epsfig{file=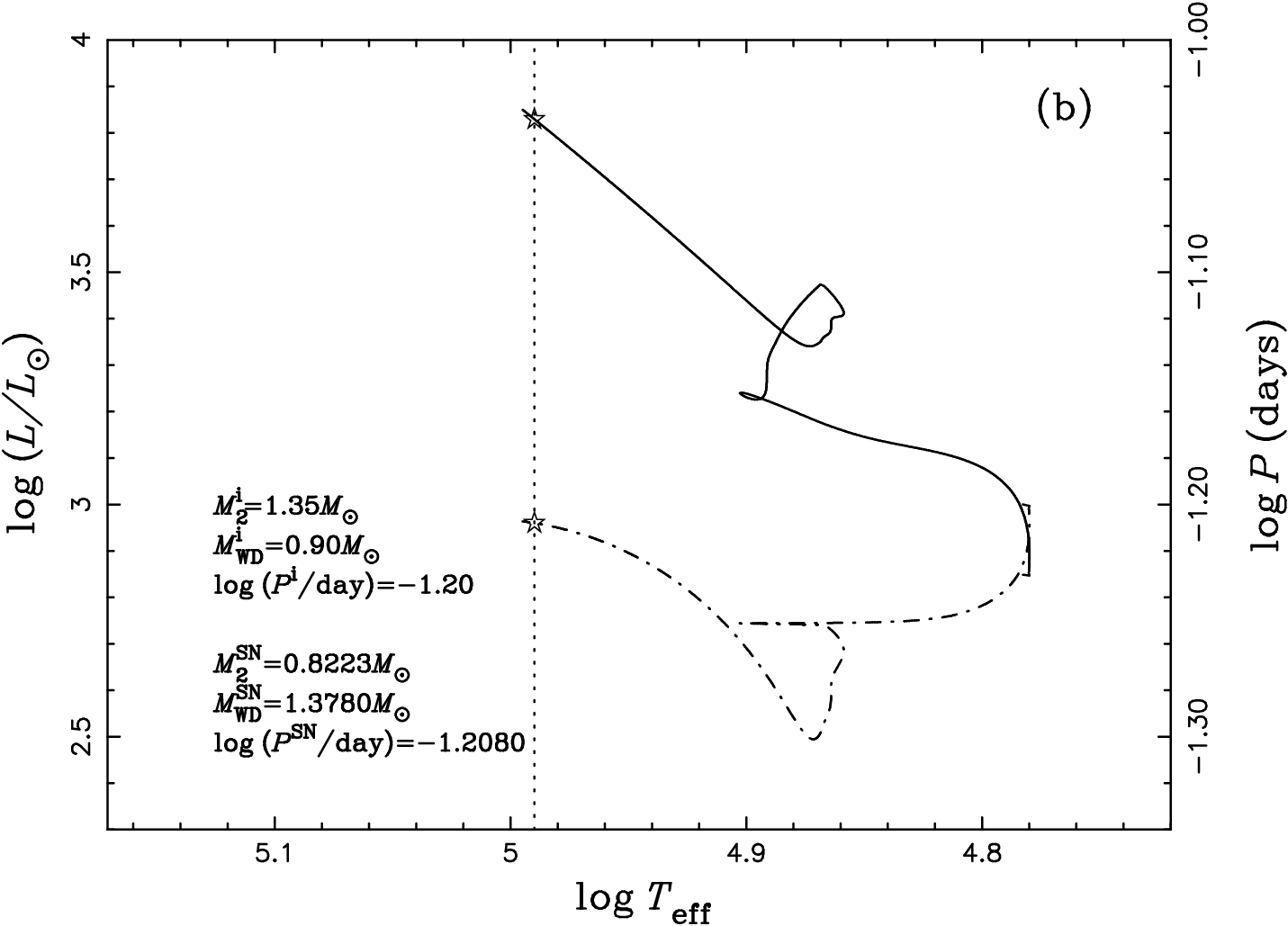,angle=270,width=8cm}} \caption{A
representative case of binary evolution calculations with $Z=0.004$,
in which the binary system is in the weak He-shell flash phase at
the moment of the SN explosion. In panel (a), the solid, dashed and
dash-dotted curves show $\dot M_2$, $\dot M_{\rm CO}$ and $M_{\rm
WD}$ varying with time, respectively. In panel (b), the evolutionary
track of the He donor star is shown as a solid curve and the
evolution of the orbital period is shown as a dash-dotted curve.
Dotted vertical lines in both panels and asterisks in panel (b)
indicate the position where the WD is expected to explode as an SN
Ia. The initial binary parameters and the parameters at the moment
of SN explosion are also given in these two panels.}
\end{figure*}

In Fig. 1, we present an example of binary evolution calculations
with $Z=0.004$. Panel (a) shows the $\dot M_2$, $\dot M_{\rm CO}$
and $M_{\rm WD}$ varying with time, while panel (b) is the
evolutionary track of the He donor star in the Hertzsprung-Russell
diagram, where the evolution of the orbital period is also shown.
The binary is ($M_2^{\rm i}$, $M_{\rm WD}^{\rm i}$, $\log (P^{\rm
i}/{\rm day})$) $=$ (1.35, 0.9, $-$1.20), where $M_2^{\rm i}$,
$M_{\rm WD}^{\rm i}$ and $P^{\rm i}$ are the initial mass of the He
star and of the CO WD in solar masses, and the initial orbital
period in days, respectively. The He star fills its Roche lobe after
the exhaustion of central He. The mass transfer rate $|\dot{M}_{\rm
2}|$ exceeds $\dot M_{\rm cr}$ soon after the onset of RLOF,
resulting in a wind phase, where a part of the transferred mass is
blown off in the form of the optically thick wind, and the left is
accumulated onto the surface of the WD. After about
$4\times10^{4}$\,yr, $|\dot{M}_{\rm 2}|$ drops below $\dot M_{\rm
cr}$ but still higher than $\dot M_{\rm st}$. Therefore, the
optically thick wind stops and the He-shell burning is stable. With
the continuous decreasing of $|\dot{M}_{\rm 2}|$, the binary system
enters into a weak He-shell flash phase about $8\times10^{4}$\,yr
later. The WD always grows in mass until it explodes as an SN Ia in
the weak He-shell flash phase. At this moment, the binary parameters
are $M^{\rm SN}_2=0.8223\,M_{\odot}$ and $\log (P^{\rm SN}/{\rm
day})=-1.2080$.

\subsection{Initial parameters for SN Ia progenitors}
\begin{figure}
\includegraphics[width=6.cm,angle=270]{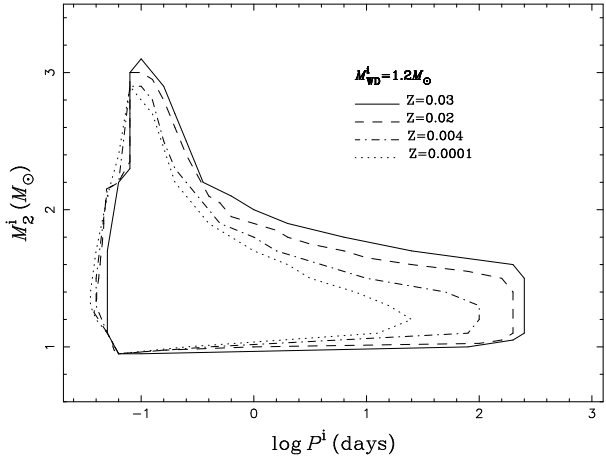}
 \caption{Parameter regions producing SNe
Ia with different metallicities in the initial orbital
period--secondary mass ($\log P^{\rm i}$, $M^{\rm i}_2$) plane of
the CO WD + He star system for initial WD mass of 1.2\,$M_{\odot}$.}
\end{figure}

Figures 2 and 3 show the initial contours for producing SNe Ia with
different metallicities, from which we can see the strong influence
of metallicity on the contours. With the increase of Z, the contours
are shifted to higher periods and larger masses, indicating that the
progenitor systems have longer periods and more massive companions
for a higher Z. This is due to the correlation between the stellar
structure and metallicity. (1) High metallicity leads to larger
radii of He ZAMS stars, so the left boundaries of the regions will
be shifted to longer period. (2) Stars with high metallicity evolve
in a way similar to those with low metallicity but less mass (Umeda
et al. 1999; Chen \& Tout 2007), resulting that, for the WD binary
systems with particular orbital periods, the companion mass
increases with metallicity. Note that, there are some dent at the
upper left boundary lines in Figs. 2 and 3. This is constrained
mainly by a high mass transfer rate because of orbit decay induced
by GWR and a large mass-ratio, leading to much of the mass being
lost from the systems in the form of the optically thick wind.

\begin{figure*}
\centerline{\epsfig{file=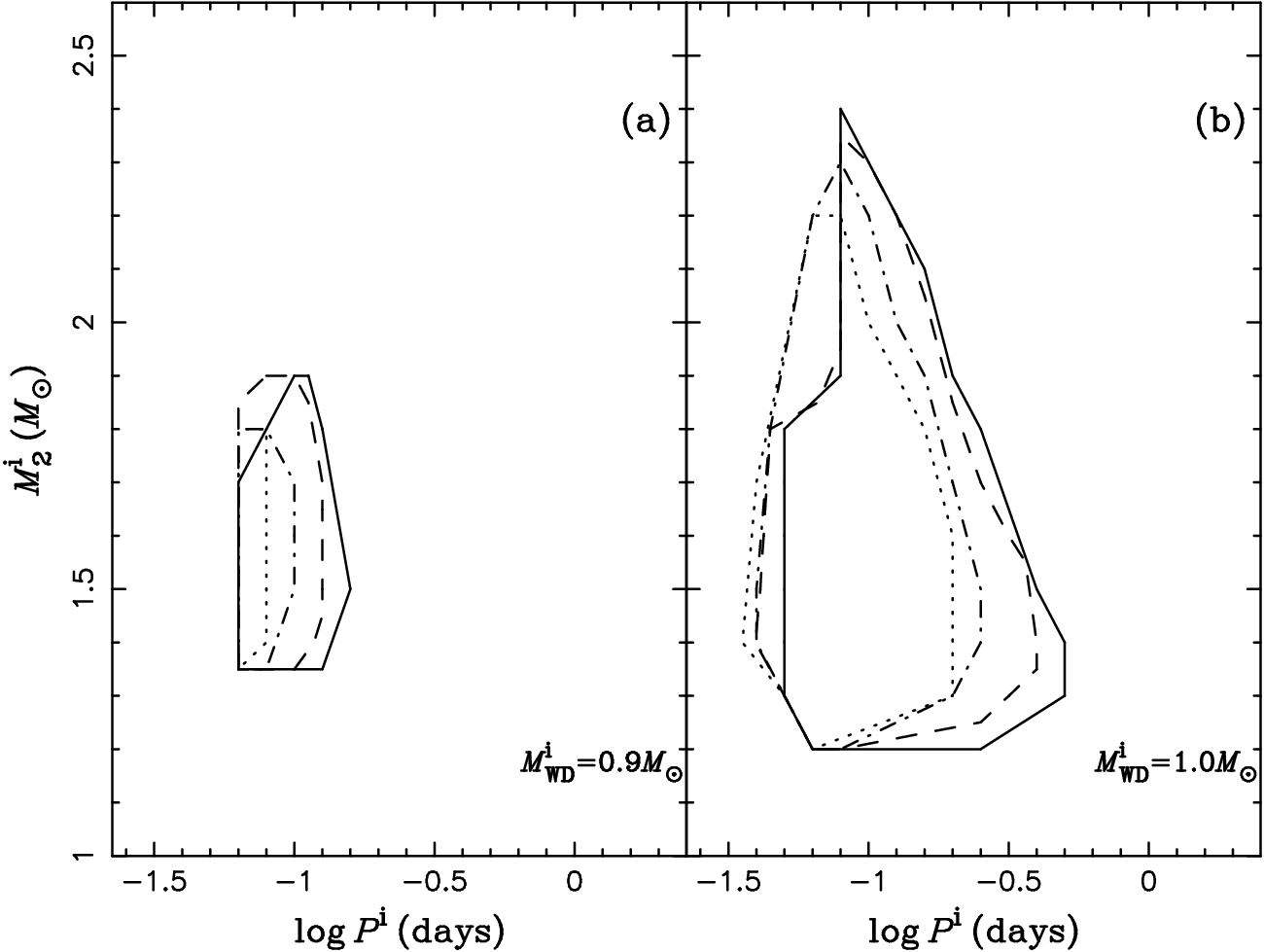,angle=270,width=8cm}\ \
\epsfig{file=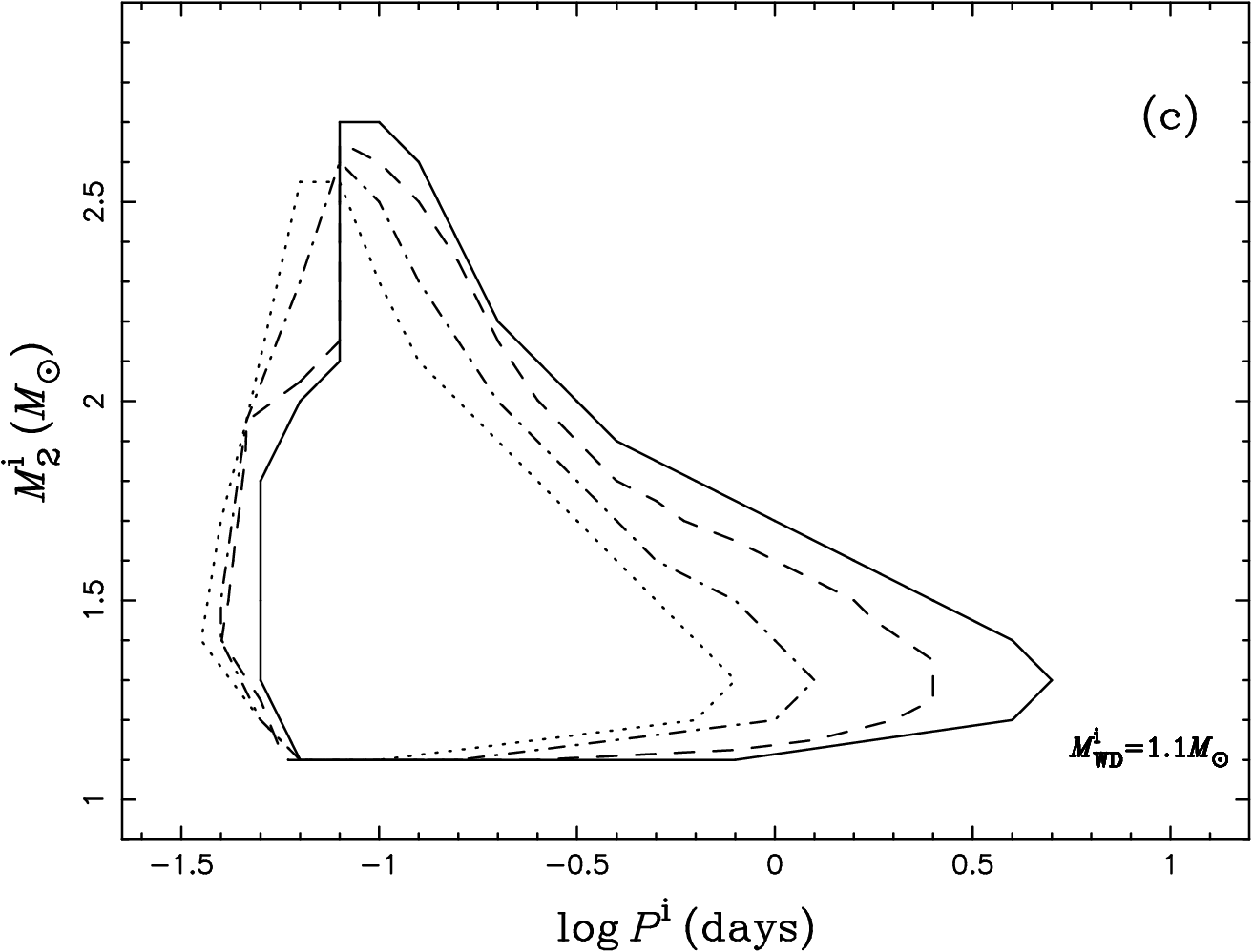,angle=270,width=8cm}} \caption{Similar to
Fig. 2, but for initial WD masses of 0.9, 1.0 and 1.1$\,M_{\odot}$.}
\end{figure*}

We also find that the minimum initial mass of the CO WDs producing
SNe Ia decreases with metallicity (e.g., for $Z=0.0001$, 0.004, 0.02
and 0.03, the minimum initial WD masses are 0.88, 0.87, 0.865 and
0.86$\,M_{\odot}$, respectively). For a high $Z$, the companions in
the WD binary systems producing SNe Ia are more massive, so more
material from the companions will be transferred onto the surface of
the WDs. Thus, the WDs do not need to be massive enough for the
production of SNe Ia, resulting in a low minimum initial WD mass.
These contours with various metallicities can be expediently used in
BPS studies. The data points and the interpolation FORTRAN code for
these contours can be supplied on request by contacting BW.

\section{Binary population synthesis}
In order to investigate SN Ia birthrates and delay times for the He
star donor channel at various metallicities, we performed a series
of Monte Carlo simulations in the BPS study. In each simulation, by
using the Hurley's rapid binary evolution code (Hurley et al. 2000,
2002), we followed the evolution of $4\times10^{\rm 7}$ sample
binaries from the star formation to the formation of the WD + He
star systems according to three evolutionary channels (i.e., the He
star channel, the EAGB channel, and the TPAGB channel; for details
see Wang et al. 2009b). We assumed that, if the initial parameters
of a CO WD + He star system at the onset of the RLOF are located in
the SN Ia production regions with a specific metallicity, an SN Ia
is produced.

\subsection{Common envelope in binary evolution}
In the He star donor channel, the progenitor of an SN Ia is a close
WD + He star system, which has most likely emerged from the common
envelope (CE) evolution of a binary involving a giant star. The CE
ejection is still an open problem. Similar to the work of Wang et
al. (2009b), we also use the standard energy equations (Webbink
1984) to calculate the output of the CE phase. For this prescription
of the CE ejection, there are two highly uncertain parameters, i.e.,
$\alpha_{\rm ce}$ and $\lambda$, where $\alpha_{\rm ce}$ is the CE
ejection efficiency, and $\lambda$ is a structure parameter that
depends on the evolutionary stage of the donor. As in previous
studies, we combine $\alpha_{\rm ce}$ and $\lambda$ into one free
parameter $\alpha_{\rm ce}\lambda$, and set it to be 1.5 which is
our standard model for the formation of the WD + He systems (Hurley
et al. 2002).

\subsection{Basic parameters for Monte Carlo simulations}
In the BPS study, the Monte Carlo simulation requires as input the
initial mass function (IMF) of the primary, the mass-ratio
distribution, the distribution of initial orbital separations, the
eccentricity distribution of binary orbit, and the star formation
rate (SFR) (e.g., Han et al. 1995a, 2002, 2003; Wang \& Han 2009,
2010c).

(1) The IMF of Miller \& Scalo (1979) is adopted. The primordial
primary is generated according to the formula of Eggleton et al.
(1989).

(2) The initial mass-ratio distribution of the binaries, $q'$, is
quite uncertain for binary evolution. For simplicity, we take a
constant mass-ratio distribution (Mazeh et al. 1992; Goldberg \&
Mazeh 1994),
\begin{equation}
n(q')=1, \hspace{2.cm} 0<q'\leq1,
\end{equation}
where $q'=M_{\rm 2}^{\rm p}/M_{\rm 1}^{\rm p}$.

(3) We assume that all stars are members of binaries and that the
distribution of separations is constant in $\log a$ for wide
binaries, where $a$ is separation and falls off smoothly at small
separation
\begin{equation}
a\cdot n(a)=\left\{
 \begin{array}{lc}
 \alpha_{\rm sep}(a/a_{\rm 0})^{\rm m}, & a\leq a_{\rm 0},\\
\alpha_{\rm sep}, & a_{\rm 0}<a<a_{\rm 1},\\
\end{array}\right.
\end{equation}
where $\alpha_{\rm sep}\approx0.07$, $a_{\rm 0}=10\,R_{\odot}$,
$a_{\rm 1}=5.75\times 10^{\rm 6}\,R_{\odot}=0.13\,{\rm pc}$ and
$m\approx1.2$. This distribution implies that the numbers of wide
binaries per logarithmic interval are equal, and that about 50\% of
stellar systems have orbital periods less than 100\,yr (Han et al.
1995b).

(4) A circular orbit is assumed for all binaries. The orbits of
semidetached binaries are generally circularized by the tidal force
on a timescale which is much smaller than the nuclear timescale.

(5) We simply assume a constant SFR over the past 15\,Gyr, or,
alternatively, a delta function, i.e., a single starburst. In the
case of the constant SFR, we calibrate the SFR by assuming that one
binary with a primary more massive than $0.8\,M_{\odot}$ is formed
annually (see Iben \& Tutukov 1984; Han et al. 1995b; Hurley et al.
2002). From this calibration, we can get ${\rm SFR}=5\,M_{\rm
\odot}{\rm yr}^{-1}$ (e.g., Willems \& Kolb 2004). For the case of
the single starburst, we assume a burst producing
$10^{11}\,M_{\odot}$ in stars. In fact, a galaxy have a complicated
star formation history. We only choose these two extremes for
simplicity. A constant SFR is similar to the situation of spiral
galaxies (Yungelson \& Livio 1998; Han $\&$ Podsiadlowski 2004),
while a delta function to that of elliptical galaxies or globular
clusters.

\section{The results of binary population synthesis}

\begin{figure}[tb]
\includegraphics[width=5.5cm,angle=270]{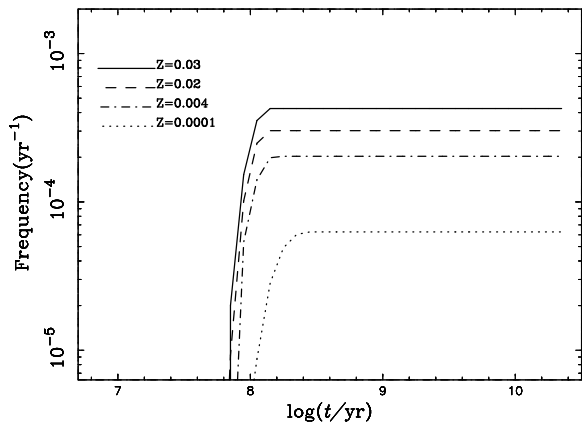}
 \caption{The evolution of SN Ia birthrates for a constant
SFR ($5\,M_{\odot}$\,yr$^{-1}$) with different metallicities. }
\end{figure}

We performed four sets of simulations with different metallicities
to systematically investigate SN Ia birthrates for the He star donor
channel. In Fig. 4, we show SN Ia birthrates for the He star donor
channel with different metallicities by adopting ${\rm
SFR}=5\,M_{\rm \odot}{\rm yr}^{-1}$. In this figure, we see that SN
Ia birthrates increase with metallicity. This is due to the fact
that the parameter space for SNe Ia increases with metallicity. The
simulations give SN Ia birthrates of $\sim$0.06$-$0.43$\times
10^{-3}\ {\rm yr}^{-1}$, which is lower than the value of
observations ($3-4\times 10^{-3}\ {\rm yr}^{-1}$ in the Galaxy; van
den Bergh \& Tammann 1991; Cappellaro \& Turatto 1997). This implies
that the He star donor channel is only a subclass of SN Ia
production, and there are some other channels or mechanisms also
contributing to SNe Ia (e.g., WD + MS channel, WD + RG channel and
double-degenerate channel). As mentioned by Wang et al. (2010), the
WD + MS channel can give a Galactic birthrate of $\sim$$1.8\times
10^{-3}\ {\rm yr}^{-1}$, and is considered to be an important
channel to produce SNe Ia.

\begin{figure}[tb]
\includegraphics[width=5.5cm,angle=270]{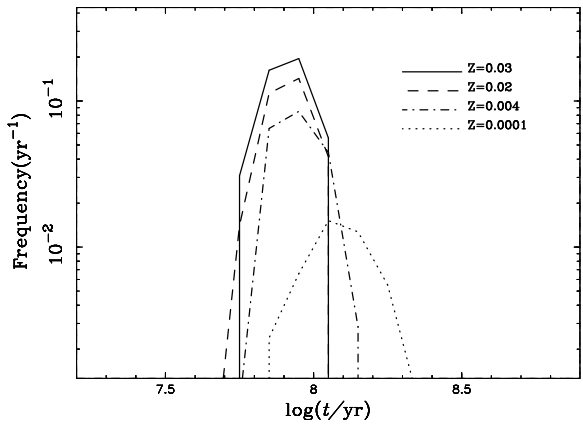}
 \caption{The evolution of SN Ia birthrates for a single starburst of $10^{\rm 11}\,M_{\odot}$ with different metallicities.}
\end{figure}

Figure 5 displays the evolution of SN Ia birthrates for a single
starburst of $10^{11}\,M_{\odot}$ with different metallicities. In
the figure SN Ia explosions occur between $\sim$45\,Myr and
$\sim$220\,Myr after the starburst, which may explain the young SNe
Ia implied by recent observations (Mannucci et al. 2006; Aubourg et
al. 2008). We see that the peak value of the birthrate is larger for
a high $Z$. This is because the parameter range of the initial WD
masses is larger for a high metallicity. We also see that a high
metallicity leads to a systematically earlier explosion time of the
SNe Ia, owing to the effects of metallicity on the maximum initial
mass of the companion for an SN Ia and on the stellar evolution. As
shown in Figs. 2 and 3, $M_{\rm 2}^{\rm i}$ increases with
metallicity. Generally, a massive star evolves more quickly than a
low-mass one. Thus, the explosion time is earlier with a high $Z$.
Although the high $Z$ also slows down the evolution of a star, its
influence is much less than that of stellar mass based on detailed
calculations of stellar evolution (Umeda et al. 1999; Chen \& Tout
2007).

The simulation in this paper was made with $\alpha_{\rm ce}\lambda
=1.5$. If we adopt a lower value for $\alpha_{\rm ce}\lambda$, e.g.,
0.5, SNe Ia occur systematically earlier for a specific metallicity.
This is because a low value of $\alpha_{\rm ce}\lambda$ tends to
have larger He star masses on average (see Fig. 5 of Wang et al.
2009b), which will evolve more quickly and hence produce an SN Ia at
an earlier time.

\section{Discussion}\label{3. Discussion}

In our binary calculations, we have not considered the influence of
rotation on the He-accreting WDs. Yoon et al. (2004) showed that, if
rotation is taken into account, He burning is much less violent than
that without rotating. This may significantly increase the
He-accretion efficiency. Also, the maximum stable mass of a rotating
WD may be above the Ch mass, i.e., the super-Ch mass WD explosions
(Uenishi et al. 2003; Yoon \& Langer 2005; Chen $\&$ Li 2009).

In addition, it is suggested that about 10\% of WDs have magnetic
fields higher than 1\,MG (Liebert et al. 2003, 2005). The mean mass
of these magnetic WDs is 0.93\,$M_{\odot}$, compared with the mean
mass (0.56\,$M_{\odot}$) of all WDs (e.g., Parthasarathy et al.
2007). Therefore, the magnetic WDs are more likely to reach the Ch
mass by accreting the He-rich material. Also, the magnetic field may
affect some properties of the WD + He star systems (e.g., the mass
transfer rate, the critical accretion rate, the thermonuclear
reaction rate, etc), resulting in a different SN Ia birthrate.

In our BPS studies, we assume that all stars are in binaries and
about 50\% of stellar systems have orbital periods less than
100\,yr. The binary fractions may depend on metallicity,
environment, spectral type, etc. If we adopt 40\% of stellar systems
have orbital periods below 100\,yr by adjusting the parameter
$a_{\rm 1}$ in equation (6), SN Ia birthrate from this channel will
decrease by 20\%. We note that the SN Ia birthrate from this channel
is low in comparison with observations, especially in the low
metallicity environment. The smaller contribution of this channel to
total SNe Ia does not change the statistics of birthrates, delay
time distribution, etc. However, the He star donor channel should
not be ignored when studying the progenitors of SNe Ia, because this
channel is considered as a main contributor to the young population
of SNe Ia (e.g., Wang \& Han 2010b; Meng \& Yang 2010). Moreover,
SNe Ia from this channel can neatly avoid H lines. We also note that
recent studies on He accretion to CO WDs reveal that
double-detonation of sub-Chandrasekhar mass WDs produces an SN Ia
that is much brighter than normal SNe Ia (e.g., Fink et al. 2010).
If this type of explosion contributes to SNe Ia, the birthrate from
the He donor star channel will increase to $\sim$$10^{-3}\ {\rm
yr}^{-1}$ (e.g., Yungelson 2005; Ruiter et al. 2009).

The companions in the He star donor channel would survive after SN
explosion and potentially be identifiable. Wang \& Han (2009) found
that the surviving companions have a high spatial velocity
($>$400\,km/s), which could be an alternative origin for
hypervelocity stars (HVSs), which are stars with a velocity so great
that they are able to escape the gravitational pull of the Galaxy.
Because SN Ia birthrates from the He star donor channel increase
with metallicity, HVSs from the SN explosion scenario are more
likely discovered in the high metallicity environment.

Currently, some observations support the existence of WD + He star
systems (e.g., KPD 1930+2752, V445 Pup, and HD 49798 with its WD
companion), which are candidates of SN Ia progenitors. (1) Maxted et
al. (2000) suggested that KPD 1930+2752 is likely to eventually
result in a merger and produce an SN Ia (see also Geier et al.
2007). However, the DD model is not supported theoretically.
Meanwhile, KPD 1930+2752 may also produce an SN Ia via the SD model,
but the parameters of the binary system are not located in the
contours of the He donor star channel for producing SNe Ia, i.e.,
KPD 1930+2752 will not produce an SN Ia via the SD model. (2) V445
Pup is an He nova (Ashok \& Banerjee 2003; Kato \& Hachisu 2003).
Kato et al. (2008) presented a free-free emission dominated light
curve model of V445 Pup, based on the optically thick wind theory
(Kato \& Hachisu 1994; Hachisu et al. 1996). The light curve fitting
in their study shows that the mass of the WD is more than
$1.35\,M_{\odot}$, and half of the accreted matter remains on the
WD, leading to the mass increase of the WD. Thus, Kato et al. (2008)
suggested that V445 Pup is a strong candidate of SN Ia progenitors
(see also Woudt et al. 2009). However, we still do not know the
orbital period of the binary system and the mass of the He donor
star so far. This needs further observations of V445 Pup after the
dense dust shell disappears. (3) HD 49798 is a H depleted subdwarf
O6 star and also a single-component spectroscopic binary with an
orbital period of 1.548\,d (Thackeray 1970; Stickland \& Lloyd
1994), which contains a X-ray pulsating companion (RX J0648.0-4418;
Israel et al. 1995). The X-ray pulsating companion is suggested to
be a massive WD (Bisscheroux et al. 1997). Based on the pulse time
delays and the binary system's inclination, constrained by the
duration of the X-ray eclipse, Mereghetti et al. (2009) recently
derived the masses of the two components. The corresponding masses
are 1.50$\pm$0.05$\,M_{\odot}$ for HD 49798 and
1.28$\pm$0.05$\,M_{\odot}$ for the WD. According to our binary
evolution model, we found the massive WD can increase its mass to
the Ch mass in future evolution. Thus, HD 49798 with its WD
companion is a likely candidate of SN Ia progenitors.

The He star donor channel with different metallicities can produce
the young SNe Ia with delay times $\sim$45$-$220\,Myr. The young
population of SNe Ia may have an effect on models of galactic
chemical evolution, since they would return large amounts of iron to
the interstellar medium earlier than previously thought. Especially,
the high metallicity environments are much earlier to return iron to
the interstellar medium, as SNe Ia from the He star donor channel
occur systemically earlier for a high $Z$. In future investigations,
we will explore the detailed influence of the young SNe Ia with
different metallicity environments on the chemical evolution of
stellar populations.

\begin{acknowledgements}
We thank the anonymous referee for his/her valuable comments that
helped us to improve the paper. This work is supported by the
National Natural Science Foundation of China (Grant No. 10821061),
the National Basic Research Program of China (Grant No.
2007CB815406), and the Chinese Academy of Sciences (Grant No.
KJCX2-YW-T24).

\end{acknowledgements}

\clearpage

\end{document}